\documentclass[10pt]{article}
\usepackage{amsmath,amsthm,natbib}
\usepackage[pdftex]{graphicx}
\usepackage{amssymb, amsfonts}
\usepackage[english]{babel}
\usepackage[ansinew]{inputenc}
\newtheorem{theorem}{Theorem}[section]

\newtheorem{proposition}[theorem]{Proposition}

\newcommand{\cov}{\operatornamewithlimits{cov}}
\newcommand{\var}{\operatornamewithlimits{var}}

\begin{document}
\title{A Loan Portfolio Model Subject to Random Liabilities and Systemic Jump Risk}
\author{Luis H. R. Alvarez\thanks{Turku School of Economics, Department of Accounting and Finance,
FIN-20500 Turku, Finland, e-mail: luis.alvarez@tse.fi} \; and \; Jani T. Sainio\thanks{Turku School of Economics, Department of Accounting and Finance,
FIN-20500 Turku, Finland, e-mail: Jani.T.Sainio@tse.fi}} \maketitle

\abstract{We extend the Vasi{\v c}ek loan portfolio model to a setting where liabilities fluctuate randomly and asset values may be subject to systemic jump risk. We derive the probability distribution of the percentage loss of a uniform portfolio and analyze its properties. We find that
the impact of liability risk is ambiguous and depends on the correlation between the continuous aggregate factor and the asset-liability ratio as well as on the default intensity.
We also find that systemic jump risk has a significant impact on the upper percentiles of the loss distribution and, therefore, on both the VaR-measure as well as on the expected shortfall.}

\thispagestyle{empty}
\clearpage
\setcounter{page}{1}

\section{Introduction}

Modeling correlation of defaults plays naturally a central role in the literature on the management of loan portfolios and valuation of credit derivatives. The reason for focusing on modeling the correlation of defaults is obvious from a credit risk management perspective: since default probabilities are relatively low, the (tail) dependence of different risks may play a dominant role in the joint probability of defaults. The role of this dependence is naturally pronounced during financial crises and, therefore, incorporating a factor (or factors) taking into account such relatively rare phenomena is clearly of interest.

In light of the fact that financial crises occur regularly and share several similar characteristics both in their precedents as well as in their negative impact on the overall economy (for excellent recent empirical studies on financial crises see Reinhart and Rogoff (2008a, 2008b, 2009)), we extend the classical  Vasi{\v c}ek loan portfolio model to a setting where the liabilities are subject to random fluctuations and asset values may face systematic asymmetric jump risk.
We follow a factorized approach (for a well-written introduction of this credit risk modeling approach, see \citet{Sch}) and model the liabilities as geometric Brownian motions driven by two continuous factors; one aggregate level factor affecting all assets and liabilities and one statistically independent idiosyncratic factor. In this way liabilities
become conditionally independent of each other given the aggregate factor dynamics. On the other hand, we assume that the value of the assets constitute exponential jump diffusions
potentially driven by three factors. As in the case of liabilities, the continuous part of the fluctuations are determined by an aggregate and an idiosyncratic factor. In addition to these factors, the assets may also be subjected to a discontinuous and spectrally negative risk component modeled as a compensated Poisson process with only positive jumps. This risk factor is assumed to be systemic in the sense that it affects all assets and, once realized, may significantly decrease their values (see, for example, \citet{EbMa} for a recent study where assets and liabilities are modeled as exponential Levy processes).  This assumption allows the analysis of rare but potentially dramatic collapses in the values of the assets backing up liabilities. Instead of considering valuation issues, we focus on the determination of the probability distribution of the percentage loss of a uniform portfolio and investigate how the interplay between different risk components affect this distribution.

It is worth emphasizing that our model is related to the pioneering work by \citet{Zh} and \citet{Zh1}. \citet{Zh} (see also \citet{Zh1}; for a recent application of Zhou's model within consumer credit setting, see \citet{AnTh}) presents a factorized model of credit risk where the value of the shares of the firm are assumed to evolve according to a geometric jump diffusion and the threshold value increases at a known constant exponential rate. Thus, liability risk is not considered as in our study. Moreover, in both \citet{Zh} and \citet{Zh1} the jumps in the value of shares may have both signs implying that even though the firm is subject to sudden large drops, it may also face unexpected significant increases in its value. In this way, his model does not take into account the realization of rare but potentially significant negative outcomes eroding the value of assets. However, in contrast to our study, \citet{Zh} derives also the arbitrage free bond prices both in the case where default occurs whenever the value falls short the known threshold value at a given fixed maturity as well as in the technically demanding first passage time setting where default occurs whenever the value falls short the threshold value prior expiry. He also studies the term structure of credit spreads and demonstrates that the considered class of models can generate a variety of yield spread curves as well as marginal default rate curves.

Our findings indicate that the impact of liability risk on the distribution of the percentage loss is ambiguous and depends, among others, on the correlation between the continuous aggregate factor and the asset-liability ratio as well as on the default intensity. If liabilities are subject to purely idiosyncratic risk and are unaffected by the aggregate market factor, then the resulting probability distribution is similar but yet not identical with the limiting distribution in the standard Vasi{\v c}ek setting. The main reason for this is that in the present setting also the idiosyncratic risk factor affects the total volatility of the percentage growth rate of the asset-liability ratio. However, in contrast with the standard  Vasi{\v c}ek model, our results indicate that there are circumstances under which the percentage loss converges to the known default probability of an individual loan. Such a case arises when the continuous aggregate factor affects equally strong both assets and liabilities. Such a case might potentially appear in situations where both assets and liabilities depend on a well diversified portfolio (for example, in the case of unit linked products).
The impact of the systemic risk component is more pronounced and our findings show that its presence may have a radical impact on the limiting distribution. First, the limiting probability distribution may have more than two modes; a phenomenon which does not arise in the standard continuous setting. Second, the systemic risk term has a significant impact on the tail probabilities and tends to increase both the upper percentiles as well as the expected shortfall associated to these percentiles even when the realization intensity is low.

The contents of this study are as follows. In section two we present the basic continuous model and state our main findings on the probability distribution of the percentage loss of a uniform portfolio. In section three we introduce the discontinuous systemic risk component and analyze its impact on the probability distribution of the percentage loss. Finally, section four concludes our study.

\section{The Impact of Liability Risk}

Our main objective is to investigate the percentage loss distribution of a loan portfolio within a conditionally independent factor model along the lines indicated by the pioneering work in \citet{Va1} (see also \citet{Va2} and \citet{Va3}). To this end, we first assume that the asset values evolve according to the random dynamics characterized by the stochastic differential equation
\begin{eqnarray}
dA_{it}=\mu_i A_{it}dt+\sigma_i A_{it}dW_{it},\quad A_{i0}=A_i,i=1,\dots,n,
\end{eqnarray}
where both the drift coefficient $\mu_i$ as well as the volatility coefficient $\sigma_i$ are exogenously given and $W_{it}$ is standard Brownian motion.
In order to model the statistical dependence of the various asset values, we assume that the driving Brownian motions can be decomposed into the form
\begin{eqnarray}
W_{it} = \sqrt{\rho_i}\;Y_t + \sqrt{1-\rho_i}\;X_{it},
\end{eqnarray}
where $Y_t, X_{1t},\dots,X_{nt}$ are a family of independent driving Brownian motions and $\rho_i\in [0,1], i=1,\dots,n$ measures the correlation between the underlying driving factor dynamics.
The factor $Y_t$ is a {\em joint aggregate risk factor} (market risk) affecting all the driving processes and the $X_{it}$'s are {\em idiosyncratic risk factors} associated to the particular asset value.

The basic Vasicek loan portfolio model assumes that the liabilities of the company are constant. However, this assumption is not always satisfied and liabilities may actually depend on the aggregate risk factor through the investment policy of the corporation (for an approach based on exponentially increasing but deterministic liabilities see \citet{Zh1}). Such a circumstance arises quite naturally, for example, in the case of unit linked insurance contracts. In order to introduce liability risk, we assume that the liabilities $B_{it}$ evolve according to the random dynamics characterized by the stochastic differential equation
\begin{eqnarray}
dB_{it}=\alpha_i B_{it}dt+\beta_i B_{it}(\sqrt{\theta_i}dY_t + \sqrt{1-\theta_i}dZ_{it}),\quad B_{i0}=B_i,
\end{eqnarray}
where both the drift coefficient $\alpha_i$ as well as the volatility coefficient $\beta_i$ are exogenously given, $\theta_i\in [0,1]$ is a coefficient measuring correlation between different liabilities, and $Z_{1t},\dots,Z_{nt}$ are a family of independent driving Brownian motions independent of the aggregate risk factor $Y_t$ and the asset-specific idiosyncratic risks $X_{1t},\dots,X_{nt}$. Thus, the liabilities are assumed to fluctuate in a similar, yet not necessarily identical, fashion with the assets.

As usually, we assume that default occurs whenever the assets do not meet the liabilities at a given date $T$. Since both the asset values as well as the liabilities follow two ordinary potentially correlated geometric Brownian motions, a standard application of It{\^o}'s lemma yields
\begin{eqnarray*}
\mathbb{P}\left[A_{iT}\leq B_{iT}\right] = \mathbb{P}\left[\frac{A_{iT}}{B_{iT}}\leq 1\right] = \Phi\left(\frac{\Xi_i}{\Sigma_i\sqrt{T}}\right),
\end{eqnarray*}
where $\Sigma_i^2 = \sigma_i^2+\beta_i^2-2\sigma_i\beta_i\sqrt{\rho_i\theta_i}$ measures the variance of the difference of the driving factors and
$$
\Xi_i = \ln\left(\frac{B_i}{A_i}\right)-\left(\mu_i-\alpha_i-\frac{1}{2}\left(\sigma_i^2-
\beta_i^2\right)\right)T.
$$

For simplicity, we assume that the recovery rate from defaulted loans in the portfolio is zero. Under this assumption, the loss $L_i$ of the $i$th loan can be defined as the random variable
$$
L_i = 1_{(0,1)}(A_{iT}/B_{iT})=\begin{cases}
1 &\textrm{ if default occurs}\\
0 &\textrm{ otherwise}
\end{cases}
$$
and, therefore, the loan portfolio percentage loss can be written as
$$
L = \frac{1}{n}\sum_{i=1}^{n}L_i.
$$
Straightforward computation shows that the probability of default conditional on the aggregate factor $Y$ now reads as
\begin{eqnarray*}
p_i(Y) =  \mathbb{P}\left[L_i=1|Y\right] = \mathbb{P}\left[\frac{A_{iT}}{B_{iT}}<1\Big{|}Y\right]
=\Phi\left(\frac{\Xi_i}{\zeta_i\sqrt{T}} - \frac{\Lambda_i}{\zeta_i}Y\right),
\end{eqnarray*}
where
$$
\zeta_i^2=\sigma_i^2(1-\rho_i)+\beta_i^2(1-\theta_i)
$$
measures the variance of the difference of the idiosyncratic risk factors, and $\Lambda_i = \sigma_i\sqrt{\rho_i}-\beta_i\sqrt{\theta_i}$
denotes the volatility multiplier of the aggregate factor $Y$ in the dynamics of the asset-liability ratio $A_{iT}/B_{iT}$.
In contrast to the standard Vasicek loan portfolio model subject to deterministically evolving liabilities, we now observe that the losses given default are
independent random variables whenever the volatility multiplier $\Lambda_i$ is identically zero for all the loans in the portfolio. As intuitively is clear, that case arises when
the aggregate factor dynamics affects both assets as well as liabilities in a similar fashion. Otherwise, the losses are statistically dependent
due to the joint dependence on the aggregate market factor. Moreover, applying the law of total probability shows that
$$
\mathbb{E}[L_iL_j] = \int_{-\infty}^{\infty}\mathbb{E}[L_iL_j|Y_T]\mathbb{P}[Y_T\in dy] = \int_{-\infty}^{\infty}p_i(\sqrt{T}y)p_j(\sqrt{T}y)\Phi'(y)dy
$$
implying that the covariance of the loss given default reads as
$$
\cov[L_i, L_j] = \int_{-\infty}^{\infty}p_i(\sqrt{T}y)p_j(\sqrt{T}y)\Phi'(y)dy - \Phi\left(\frac{\Xi_i}{\Sigma_i\sqrt{T}}\right)\Phi\left(\frac{\Xi_j}{\Sigma_j\sqrt{T}}\right).
$$
Along the lines of our observations above, we find that if $\Lambda_i=\Lambda_j=0$ then $\cov[L_i,L_j] =0$. It is worth emphasizing that
these covariances (and, therefore, default correlations) are typically very sensitive with respect to changes in the maturity $T$ of the loans.

In order to investigate the probability distribution of the percentage loss of a loan portfolio, let us now assume that the portfolio is formed by $n$  identical contracts and denote the unconditional probability of default of an individual loan as
$p$. In that case we observe that the probability of default conditional on the aggregate factor $Y$ can be expressed as
\begin{eqnarray}
p(Y) = \Phi\left(\frac{1}{\zeta}\left(\Sigma\Phi^{-1}(p)-\Lambda Y\right)\right).\label{condprob1}
\end{eqnarray}
Given this expression, denote now as $\bar{L}=\lim_{n\rightarrow\infty}L$ the limiting loan portfolio percentage loss of a infinitely large portfolio. We can now establish the following:
\begin{proposition}\label{prop1}
The probability of $k$ defaults in the loan portfolio percentage loss reads as
\begin{eqnarray}
\mathbb{P}\left[L=\frac{k}{n}\right] = {n\choose k}\int_{-\infty}^{\infty}p^k(\sqrt{T}y) (1-p(\sqrt{T}y))^{n-k}\Phi'(y)dy,\label{binomial}
\end{eqnarray}
where $p(y)$ is given in \eqref{condprob1}. If $\Lambda = 0$ then loan portfolio percentage loss converges almost certainly to the deterministic limit $\bar{L}=p$. However, if $\Lambda\neq 0$ then the limiting loan portfolio percentage loss is distributed according to the probability distribution
\begin{eqnarray}
\mathbb{P}[\bar{L}\leq x] = \Phi\left(\frac{1}{|\Lambda|}\left(\zeta \Phi^{-1}(x) - \Sigma \Phi^{-1}(p)\right)\right)\label{limit1}
\end{eqnarray}
with density
\begin{eqnarray}
f(x) = \frac{\zeta}{|\Lambda|}\frac{\Phi'\left(\frac{1}{|\Lambda|}\left(\zeta \Phi^{-1}(x) - \Sigma \Phi^{-1}(p)\right)\right)}{\Phi'(\Phi^{-1}(x))}.\label{density1}
\end{eqnarray}
\end{proposition}
\begin{proof}
The binomial formula \eqref{binomial} is a direct implication of the law of total probability and the binomial nature of the loss given default (see, for example, chapter 9 in \citet{La} and chapter 8 in \citet{McFrEm}). On the other hand, since the losses given default are conditionally independent, we observe that the conditions of the strong law of large numbers (SLLN) are satisfied and, therefore, that the percentage loss conditional on the aggregate factor converges to its expectation which, in the present case, reads as in \eqref{condprob1} when $\Lambda\neq 0$ and as $p$ when $\Lambda=0$.  Equation \eqref{limit1} then follows by computing the probability $\mathbb{P}[p(Y)\leq x]$. The density can then be derived by ordinary differentiation.
\end{proof}

Proposition \ref{prop1} extends the results of the standard Vasicek loan portfolio model to the case where also liabilities are subject to random fluctuations.
The main difference with the standard model is that now the volatility multiplier of the aggregate market factor in the dynamics of the asset-liability ratio can be zero even in the case where the factor affects both assets as well as liabilities. If this multiplier is zero, then the losses are IID random variables and the probability distribution can be directly analyzed in terms of constant binomial probabilities. In that case, the percentage loss converges almost everywhere to the known binomial probability. However, if the multiplier is not zero, then the percentage loss converges towards a random variable with known distribution \eqref{limit1} which resembles, but is not identical, with the limiting distribution in the case of constant liabilities.

\begin{figure}[h!]
\begin{center}
\includegraphics[width=0.7\textwidth]{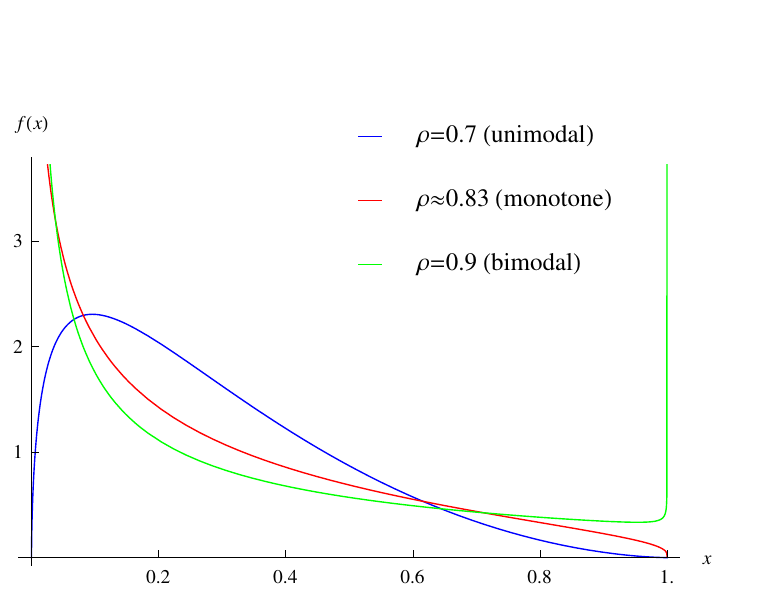}
\end{center}
\caption{\small Loss densities}\label{ratiofig3}
\end{figure}

Straightforward computations show that the probability density function is bimodal when $\Lambda^2 > \zeta^2$, monotone when $\Lambda^2 = \zeta^2$, and unimodal with mode at
$$
\bar{L}_M = \Phi\left(\frac{\zeta \Sigma}{\zeta^2-\Lambda^2}\Phi^{-1}(p)\right)
$$
when $\Lambda^2 < \zeta^2$. Consequently, along the original observations by Vasicek we find that depending on the precise parametrization of the model, the distribution may be either unimodal or bimodal and it can also be very skewed.
We illustrate the loss density in Figure \ref{ratiofig3} for various correlations under the assumptions that $\theta=0.7$, $\sigma=0.2$, $\beta=0.1$,
$\mu=0.055$, $\alpha=0.05$, $T=1$, $B_0=1$, and $A_0=1.1$.

\begin{figure}[h!]
\begin{center}
\includegraphics[width=0.7\textwidth]{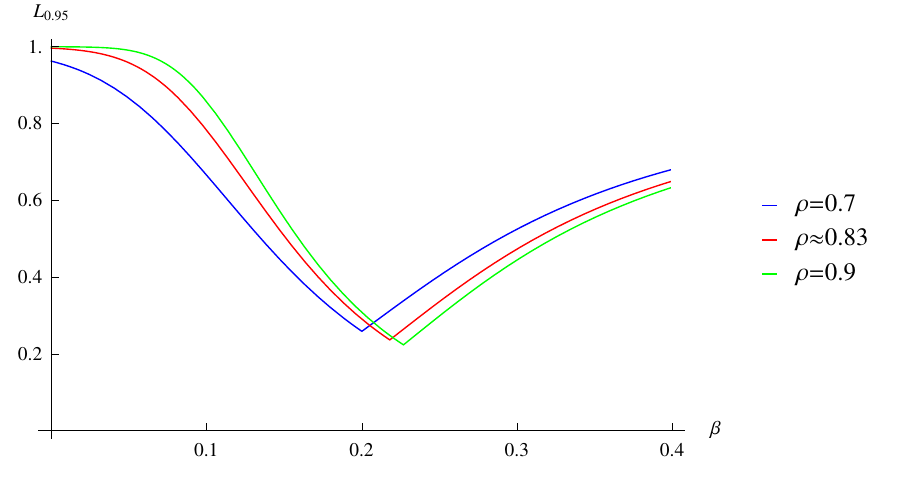}
\end{center}
\caption{\small The Impact of Liability Risk on the Percentile $L_{0.95}$}\label{ratiofig3a}
\end{figure}

In the present setting the $\nu$-percentile $L_\nu$ satisfying the identity
$\mathbb{P}[\bar{L}\leq L_\nu] =  \nu$ is
$$
L_\nu = \Phi\left(\frac{1}{\zeta}\left(\Sigma \Phi^{-1}(p)+|\Lambda|\Phi^{-1}(\nu)\right)\right).
$$
The percentile $L_\nu$ depends, among others, on the volatility $\beta$ of the liabilities. Unfortunately, it is not monotonic as a function of
$\beta$ and, therefore, the impact of liability risk on the percentiles is ambiguous. The 95\% percentile $L_{0.95}$ is illustrated as a function of $\beta$ in Figure \ref{ratiofig3a} under the assumptions
$\theta=0.7$, $\sigma=0.2$,
$\mu=0.055$, $\alpha=0.05$, $T=1$, $B_0=1$, and $A_0=1.1$.

\section{The Effect of Random Systemic Risk}

Having considered the impact of liability risk on the limiting probability distribution of the loan portfolio percentage loss, we follow the original study by \citet{Zh} (see also \citet{Zh1}) and extend our basic model to the case where the assets backing up liabilities are subject to unexpected random jumps modeled as a compound Poisson process. In contrast with \citet{Zh}, we assume that these unexpected jumps are only one-sided (downward jumps) and occur at the aggregate level. Therefore, the driving compound process is a common factor affecting all assets; an assumption permitting the analysis of the the impact of rare but potentially significant collapses (i.e. realization of systemic risk) in the asset values to the limiting default intensity in a large loan portfolio.

In line with these arguments, we now assume that the asset values evolve according to the dynamics
\begin{eqnarray}
A_{it} =  A_ie^{\left(\mu_i+\lambda (1-\mathbb{E}[e^{-\xi_1}])-\frac{1}{2}\sigma_i^2\right)t+\sigma_i (\sqrt{\rho_i}Y_t + \sqrt{1-\rho_i}X_{it})-J_t},\label{jumpdyn}
\end{eqnarray}
where
\begin{eqnarray}
J_t = \sum_{k=0}^{N_t}\xi_k\label{compPo}
\end{eqnarray}
is a compound Poisson process independent of the continuous aggregate factor $Y$. In \eqref{compPo}, we assume $N_t$ is a standard Poisson process with intensity $\lambda$, $\{\xi_k\}_{k\geq 1}$ is a sequence of nonnegative iid
random variables with known distribution, and $\xi_0=0$. In equation \eqref{jumpdyn} $\lambda (1-\mathbb{E}[e^{-\xi_1}])$ is a compensation term needed to guarantee that the asset value is expected to grow at the same rate as in the absence of jumps. If this compensation term is not taken into account then the proposed asset value model is almost surely lower and has a smaller expected value than the model considered in the previous section (due to the nonnegativity of the jumps and the monotonicity of the driving Poisson process). Especially, we observe that \eqref{jumpdyn} can be expressed as
\begin{eqnarray}
dA_{it} = \mu_i A_{it}dt + \sigma_i A_{it}dW_{it}+A_{it}\int_{\mathbb{R}}(e^{-z}-1)d\tilde{N}(dt,dz),\label{jumpdynsde}
\end{eqnarray}
where $\tilde{N}(dt,dz)$ denotes the Poisson random measure associated to the underlying compensated Poisson process (cf. Chapter 2 in \citet{Kyp1}).

It is worth pointing out that the stated specification results into an asset value which coincides in the mean but is more volatile than the model in the absence of systemic jumps. More precisely, it is clear that now that for all $t$ it holds $\mathbb{E}[A_{it}]=A_i e^{\mu_i t}$ and
$$
\var[A_{it}] = A_i^2e^{2\mu_i t}\left(e^{\sigma_i^2 t + \lambda t \mathbb{E}[(1-e^{-\xi_1})^2]}-1\right) > A_i^2e^{2\mu_i t}\left(e^{\sigma_i^2 t}-1\right).
$$
In this way the considered process can be interpreted as a {\em mean preserving spread} of the continuous asset value dynamics considered in the previous section.

Applying an analogous conditioning argument as in the previous section, we now find that the probability of default given the aggregate factors $Y$ and $J$ is
$$
\mathbb{P}[A_{iT}\leq B_{iT}|Y,J] = \Phi\left(\frac{\tilde{\Xi}}{\zeta_i\sqrt{T}}-\frac{\Lambda_i}{\zeta_i\sqrt{T}}Y_T + \frac{J_T}{\zeta_i\sqrt{T}}\right),
$$
where
$
\tilde{\Xi}_i = \Xi -\lambda (1-\mathbb{E}[e^{-\xi_1}])T.
$
As intuitively is clear, the positivity of the jump component $J_T$ implies that the probability of default is in this setting higher that in the absence of unexpected
downward jumps in the value of the assets. However, it is not beforehand clear how significant the effect of the Poisson component on the default probability is, and how
this effect depends on both the intensity of the driving Poisson process and the precise nature of the jump size distribution. Moreover, in the present setting
the loans are statistically dependent even when the volatility multiplier of the aggregate market factor in the dynamics of the asset-liability ratio is zero (i.e. $\Lambda_i=0$ for all
$i$). The reason for this is naturally the presence of the systemic jump risk component affecting all assets.

In order to be able to analyze the limiting probability distribution of the percentage portfolio loss, we now again assume
that we have a portfolio of $n$ approximately identical contracts.
In that case we find that the conditional probability of default given the aggregate factors reads as
$$
p(Y, J) = \Phi\left(\frac{\Sigma}{\zeta}\left(\Phi^{-1}(\tilde{p})-\frac{\Lambda}{\Sigma\sqrt{T}}Y_T + \frac{J_T}{\Sigma\sqrt{T}}\right)\right),
$$
where
$$
\tilde{p} = \Phi\left(\frac{\Xi -\lambda (1-\mathbb{E}[e^{-\xi_1}])T}{\Sigma \sqrt{T}}\right).
$$
We can now establish the following result:
\begin{proposition}\label{prop2}
If $\Lambda=0$ then the limiting loan portfolio percentage loss is distributed according to the probability distribution
\begin{eqnarray}
\begin{split}
\mathbb{P}[\bar{L}\leq x] = e^{-\lambda T}\chi_{[p,1]}(x) + \sum_{k=1}^{\infty}e^{-\lambda T}\frac{(\lambda T)^k}{k!}\int_0^{M_T}\mathbb{P}\left[S_k\in du\right],
\end{split}\label{limit2a}
\end{eqnarray}
where $M_T =\Sigma\sqrt{T}\left(\Phi^{-1}(x)-\Phi^{-1}(\tilde{p})\right)$ and
$$
\mathbb{P}[S_k\in du] = \mathbb{P}\left[\sum_{j=1}^k\xi_j\in du\right] = (g\ast\cdots\ast g)(u)du
$$
is the $k$-fold convolution of the density $g(u)$ of the random jump-size. If, however, $\Lambda\neq 0$ then
\begin{eqnarray}
\begin{split}
\mathbb{P}[\bar{L}\leq x] &= e^{-\lambda T}\Phi\left(H(x,0)\right) \\
&+ \sum_{k=1}^{\infty}e^{-\lambda T}\frac{(\lambda T)^k}{k!}\int_0^\infty \Phi\left(H(x,u)\right)\mathbb{P}[S_k\in du],
\end{split}\label{limit2}
\end{eqnarray}
where
\begin{eqnarray*}
H(x,u) = \frac{1}{|\Lambda|}\left(\zeta \Phi^{-1}(x) - \Sigma \Phi^{-1}(\tilde{p})-\frac{u}{\sqrt{T}}\right).
\end{eqnarray*}
In this case, the density of the loan portfolio percentage loss reads as
\begin{eqnarray}
\begin{split}
\hat{f}(x) &= \frac{\zeta}{|\Lambda|}\int_0^\infty \sum_{k=1}^{\infty}e^{-\lambda T}\frac{(\lambda T)^k}{k!}\frac{\Phi'(H(x,u))}{\Phi'(\Phi^{-1}(x))}\mathbb{P}[S_k\in du]\\
&+ e^{-\lambda T} \frac{\zeta}{|\Lambda|}\frac{\Phi'(H(x,0))}{\Phi'(\Phi^{-1}(x))}
\end{split}\label{density2}
\end{eqnarray}
\end{proposition}
\begin{proof}
As in Proposition \ref{prop1}, the losses given default are conditionally independent given the aggregate factors and satisfy the conditions of the SLLN.
The probability distributions \eqref{limit2a} and \eqref{limit2} follow directly by invoking the law of total probability in computing the probability $\mathbb{P}[p(Y, J) \leq x]$. The density \eqref{density2} can then be derived with ordinary differentiation.
\end{proof}
Proposition \ref{prop2} states the limiting probability distribution and its density for a sufficiently large loan portfolio percentage loss.
Unfortunately, the distribution is in this case very complicated (being a mixture; for a comprehensive treatment of mixtures within credit risk management applications, see Chapter 8 in \citet{McFrEm}) and identifying the percentiles explicitly is extremely demanding, if possible at all. However, it is worth emphasizing that in contrast to the case subject to continuous factor dynamics, the distribution may now be multimodal. The reason for this observation is that now
the density $\hat{f}(x)$ is a probability weighted sum of potentially bimodal densities. More precisely, since
$$
\frac{\Phi'(H(x,u))}{\Phi'(\Phi^{-1}(x))} = e^{\frac{1}{2}\Phi^{-1}(x)^2 - \frac{1}{2\Lambda^2}\left(\zeta \Phi^{-1}(x) - \Sigma \Phi^{-1}(\tilde{p})-\frac{u}{\sqrt{T}}\right)^2}
$$
is bimodal whenever $\Lambda^2>\zeta^2$, we notice that the limiting distribution may be multimodal depending on the jump size distribution. For example, when the jump size is
a known constant, the limiting distribution may have more modes than just two.

\begin{figure}[h!]
\begin{center}
\includegraphics[width=0.7\textwidth]{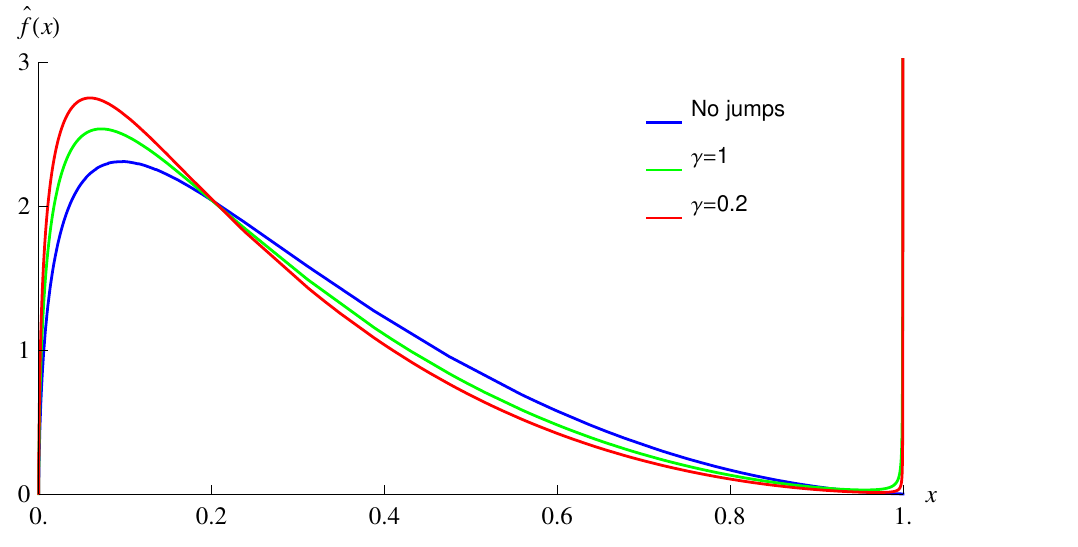}
\end{center}
\caption{\small Loss densities}\label{ratiofig4}
\end{figure}

In order to investigate numerically the impact of jumps on the limiting distribution of the loan portfolio percentage loss, we now consider the special case where
the jump size is exponentially distributed with parameter $\gamma$. It is well-known that in this case
the series
$$S_n = \sum_{k=1}^{n}\xi$$
is Gamma-distributed according to the density
$$
\mathbb{P}[S_n\in du] = \frac{\gamma e^{-\gamma u}(\gamma u)^{n-1}}{(n-1)!}du.
$$
In this case the density of the loan portfolio percentage loss reads as
\begin{eqnarray*}
\hat{f}(x) &=& \frac{\zeta}{|\Lambda|}\int_0^\infty \sum_{k=1}^{\infty}e^{-\lambda T}\frac{(\lambda T)^k}{k!}\frac{\Phi'(H(x,u))}{\Phi'(\Phi^{-1}(x))}\frac{\gamma e^{-\gamma u}(\gamma u)^{k-1}}{(k-1)!}du\\
&+& e^{-\lambda T} \frac{\zeta}{|\Lambda|}\frac{\Phi'(H(x,0))}{\Phi'(\Phi^{-1}(x))}.
\end{eqnarray*}
We illustrate this density in the three different cases arising in the absence of jump risk. Figure \ref{ratiofig4} illustrates the case where the limiting distribution is unimodal under the assumptions that $\theta=\rho=0.7$, $\sigma=0.2$, $\beta=0.1$,
$\mu=0.055$, $\alpha=0.02$, $T=1$, $\lambda=0.02$, $B_0=1$, and $A_0=1.1$.
As is clear from Figure \ref{ratiofig4}, the presence of downward jump risk has a pronounced impact on the upper tail of the limiting density and,
therefore, on the percentiles of the distribution. These percentiles are numerically illustrated in the following table.
\begin{table}[h!]
  \centering
  \begin{tabular}{|c|c|c|c|c|c|c|}
    \hline
    $\nu$                        & 0.90  & 0.915  & 0.93 & 0.945 & 0.96 & 0.975 \\
    \hline
    $\gamma \rightarrow\infty$ & 57.1&59.52&62.23&65.37&69.12&73.97 \\
    $\gamma = 1$               & 56.5&59.32&62.61&66.6&71.81&80.01 \\
    $\gamma = 0.2$             & 54.65&57.57&61.01&65.25&70.98&81.02\\
    \hline
  \end{tabular}
  \caption{{\small Percentiles in the Case of Figure \ref{ratiofig4}}} \label{tab1}
\end{table}
As Table \ref{tab1}
clearly illustrates the difference between the percentiles is significant for sufficiently high percentiles. For example, in the absence of the systemic jump component the percentage loss exceeds 73.97\% with probability 2.5\%. In the presence of the systemic jump component this percentile is radically changed and the percentage loss is expected to exceed 80.01\% (81.02\%) with the same probability. The expected shortfalls associated with the percentiles appearing on Table \ref{tab1} are illustrated on Table \ref{tab3}.
\begin{table}[h!]
  \centering
  \begin{tabular}{|c|c|c|c|c|c|c|}
    \hline
    $\nu$                       & 0.90  & 0.915  & 0.93 & 0.945 & 0.96 & 0.975 \\
    \hline
    $\gamma \rightarrow\infty$ & 68.47&70.26&72.28&74.61&77.39&80.97 \\
    $\gamma = 1$               & 72.7&75.31&78.39&82.17&87.09&94.05 \\
    $\gamma = 0.2$             & 72.35&75.22&78.65&82.9&88.52&96.43 \\
    \hline
  \end{tabular}
  \caption{{\small Expected Shortfall $ES_{\nu}$}} \label{tab3}
\end{table}
As Table \ref{tab3} shows, the impact of the systemic jump component on the expected shortfalls is significant as well. Interestingly, the difference becomes higher as the confidence limit increases. The reason for this observation is the skewness of the density towards higher realizations in the presence of the systemic jump component.

\begin{figure}[h!]
\begin{center}
\includegraphics[width=0.7\textwidth]{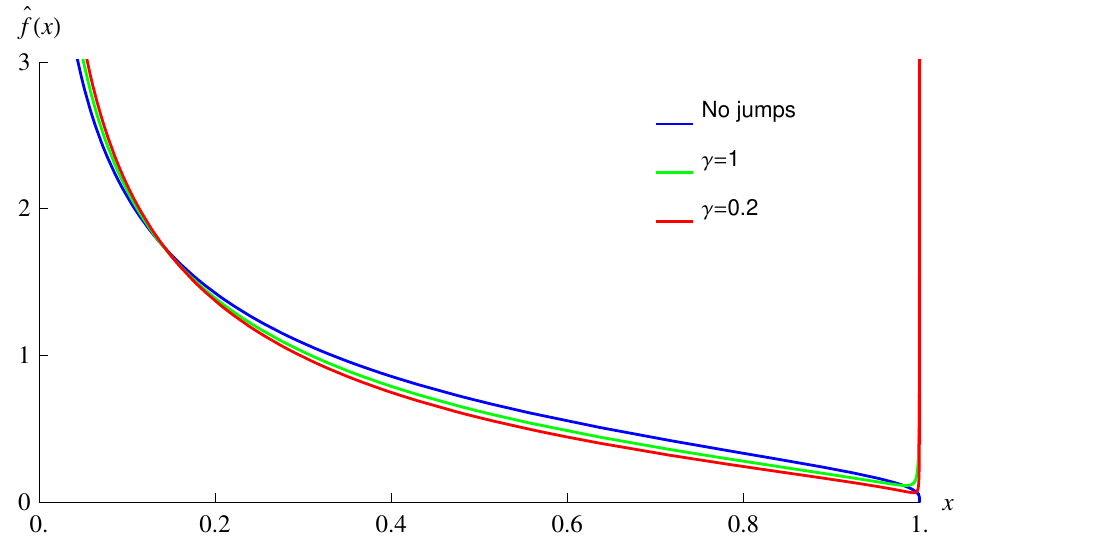}
\end{center}
\caption{\small Loss densities}\label{ratiofig6}
\end{figure}

For the sake of comparison, the case where the limiting density is monotone in the absence of jump risk is illustrated
in Figure \ref{ratiofig6}  under the assumptions that $\theta=0.7$, $\rho\approx 0.83$, $\sigma=0.2$, $\beta=0.1$,
$\mu=0.055$, $\alpha=0.05$, $T=1$, $\lambda=0.02$, $B_0=1$, and $A_0=1.1$.

\begin{table}[h!]
  \centering
  \begin{tabular}{|c|c|c|c|c|c|c|}
    \hline
    $\nu$                        & 0.90  & 0.915  & 0.93 & 0.945 & 0.96 & 0.975\\
    \hline
    $\gamma \rightarrow\infty$ & 66.17&69.42&72.96&76.85&81.23&86.34\\
    $\gamma = 1$               &  65.94&69.71&73.91&78.7&84.38&91.69\\
    $\gamma = 0.2$             & 63.91&67.89&72.37&77.59&83.96&92.84\\
    \hline
  \end{tabular}
  \caption{{\small Percentiles in the Case of Figure \ref{ratiofig6}}} \label{tab2}
\end{table}
Again we notice from Table \ref{tab2} that
the impact of the discontinuous systemic risk component on the percentiles of the percentage loss distribution is significant. For example, in the absence of the systemic jump component the percentage loss exceeds 86.34\% with probability 2.5\%. In the presence of the systemic jump component these percentile is 91.69\% (92.84\%). The expected shortfalls associated with the percentiles appearing on Table \ref{tab2} are now, in turn, illustrated on Table \ref{tab4}.
\begin{table}[h!]
  \centering
  \begin{tabular}{|c|c|c|c|c|c|c|}
    \hline
    $\nu$                        & 0.90  & 0.915  & 0.93 & 0.945 & 0.96 & 0.975\\
    \hline
    $\gamma \rightarrow\infty$ &  79.47&81.54&83.76&86.18&88.88&91.98\\
    $\gamma = 1$               &  82.26&84.81&87.61&90.7&94.18&97.98\\
    $\gamma = 0.2$             &  81.66&84.45&87.53&90.97&94.84&98.91\\
    \hline
  \end{tabular}
  \caption{{\small Expected Shortfall $ES_{\nu}$}} \label{tab4}
\end{table}

\section{Conclusions}

We considered the impact of liability risk on the percentage loss distribution of a large uniform loan portfolio both in the presence and in the absence of discontinuous systemic risk.
As our findings show, the impact of liability risk is ambiguous and it may increase or decrease the percentiles depending on the precise parametrization of the considered model and, especially, on the strength of the dependence between the asset-liability-ratio and the the driving continuous aggregate factor. The discontinuous jump factor capturing the systemic risk has a more pronounced impact on the limiting percentage loss distribution since it affects all the asset-liability-ratios through the asset values. Our results seem to indicate that its impact becomes more significant at the tails of the distribution, which are found to be bimodal  in the exponential case. According to our findings, the presence of systemic risk affects in a relatively significant way the expected shortfall associated to the upper tail probabilities even when the realization of the risk is assumed to be rare.

There are several directions towards which our model could be generalized. First, the considered loan portfolio is assumed to be large and uniform, thus overlooking the potentially significant effect of the granularity of a loan portfolio. Second, assuming that there is no recovery once default has occurred is another simplifying assumption which could relaxed. Third, our analysis focuses solely on the distribution of the percentage loss distribution and overlooks the pricing of bonds within the considered setting. All these interesting questions are left for future research.\\

\noindent{\bf Acknowledgements}: The authors are grateful to the deputy managing director at Federation of Finnish Financial Services {\em Esko Kivisaari} for proposing this research subject and to {\em Teppo Rakkolainen} for insightful comments on the contents of the study. The financial support from the {\em OP Bank Group Research Foundation} is gratefully acknowledged.

\end{document}